\documentclass[11pt,letterpaper]{article}
\usepackage{systeme} 
\usepackage[utf8]{inputenc}
\usepackage{multirow} 
\usepackage{booktabs} 
\usepackage[english]{babel}
\usepackage{amsmath}
\usepackage{amsfonts}
\usepackage{amssymb}
\usepackage{graphicx}
\usepackage[small,bf]{caption} 
\usepackage[left=3cm,right=3cm,top=2cm,bottom=3cm]{geometry}

\author{Jorge Pinochet}
\title{\textbf{Three easy ways to the Hawking temperature}}
\begin{document}

\author{Jorge Pinochet$^{*}$\\ \\
 \small{$^{*}$\textit{Departamento de Física, Universidad Metropolitana de Ciencias de la Educación,}}\\
 \small{\textit{Av. José Pedro Alessandri 774, Ñuñoa, Santiago, Chile.}}\\
 \small{e-mail: jorge.pinochet@umce.cl}\\}

\date{}
\maketitle

\begin{center}\rule{0.9\textwidth}{0.1mm} \end{center}
\begin{abstract}
\noindent In this work, three heuristic derivations of the Hawking temperature are presented. The main characteristic of these derivations is their extreme simplicity, which makes them easily accessible to a wide and diverse audience. \\ \\


\begin{center}\rule{0.9\textwidth}{0.1mm} \end{center}
\end{abstract}

\maketitle

\section{Introduction}

Stephen Hawking's most important scientific contribution was his theoretical demonstration that black holes (BHs) are not really black after all, since they produce a thermal emission, the \textit{Hawking radiation}, and therefore have a characteristic temperature known as \textit{Hawking temperature}. This discovery is too important to only remain inside the scientific community. The objective of this work is to present three simple heuristic derivations of the Hawking temperature that are accessible to a wide and diverse audience, ranging from high school students to engineers and science teachers. Each derivation takes a different physical perspective, contributing to a better understanding of Hawking's discovery, as well as its underlying physical concepts. This article also serves as a complement to other works on the subject published in this journal, which analyse the great findings of Hawking in detail [1–6].\\ 

To achieve this goal, we only address concepts that are strictly necessary to carry out the derivations. Readers who are interested in delving into the topics covered in more detail are encouraged to consult the articles cited above.

\section{BHs and quantum fluctuations}

A BH is a region of space where gravity is so strong that nothing can escape from it, not even light [7]. Figure 1 shows the internal structure of a the simplest type of BH. This object has a spherical horizon located at a distance $R_{S}$ from the central singularity [7,8], where

\begin{equation}
R_{S} = \frac{2GM_{BH}}{c^{2}},
\end{equation}

is the gravitational radius, $G = 6.67 \times 10^{-11} N\cdot m^{2} \cdot kg^{-2}$ is the gravitational constant, $c = 3 \times 10^{8} m\cdot s^{-1}$ is the speed of light in vacuum, and $M_{BH}$ is the mass of the BH.\\ 

What happens if we expand the purely gravitational description of a BH and incorporate quantum mechanics? This was the key question that led Hawking to discover that BH are not really black. In brief terms, Hawking's discovery is as follows.\\

\begin{figure}[h]
  \centering
    \includegraphics[width=0.3\textwidth]{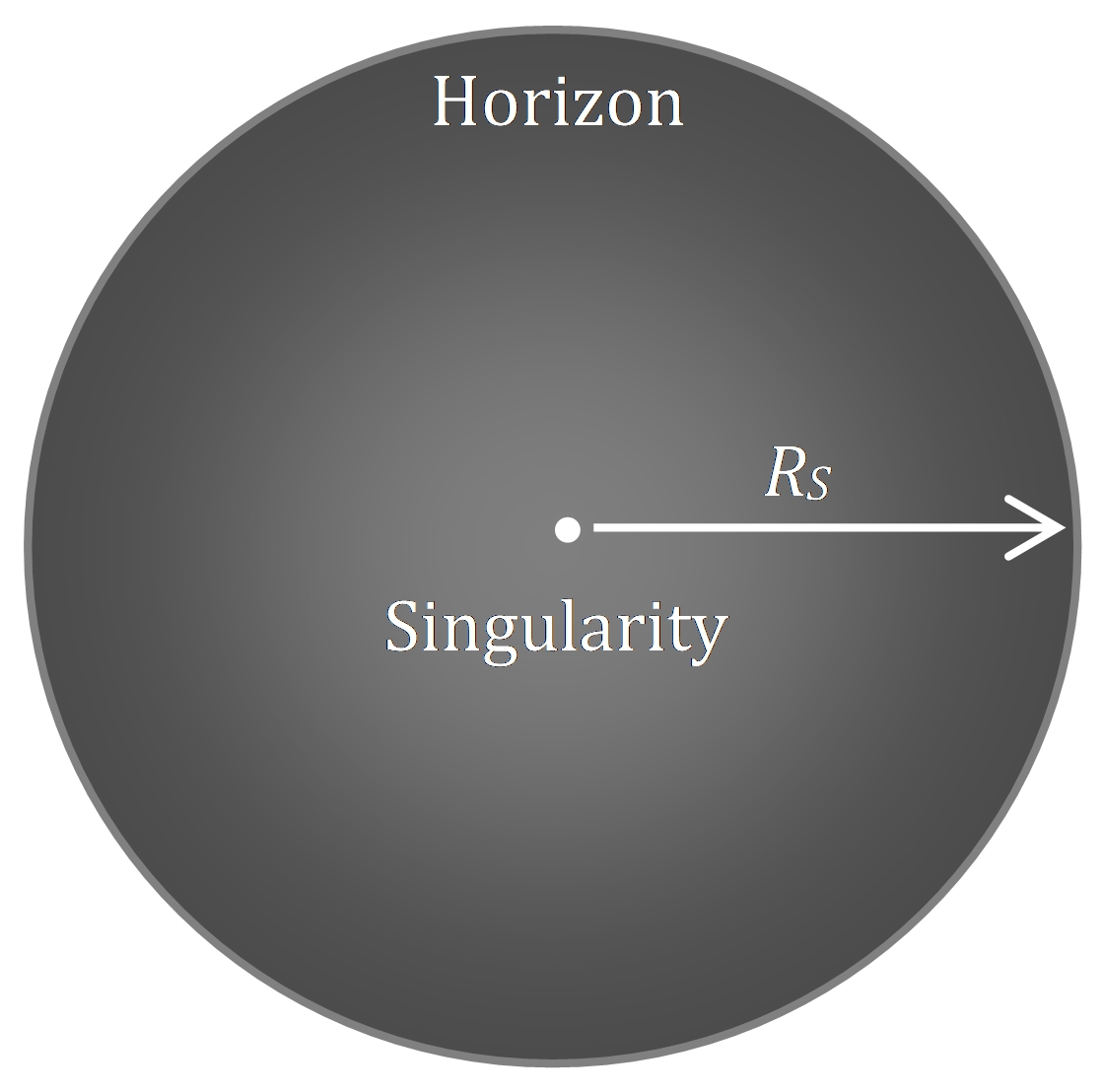}
  \caption{A static BH that is spherically symmetric and has an event horizon of radius RS and a central singularity.}
\end{figure}

According to the Heisenberg uncertainty principle, which is one of the most fundamental results of quantum mechanics, a phenomenon known as \textit{quantum fluctuation} occurs constantly and randomly at every point in space. This phenomenon consists of the appearance and disappearance of particle-antiparticle pairs in a vacuum [9]. The particles/antiparticles produced by quantum fluctuations are "virtual" because the time between their appearance and disappearance is always below the detection threshold. However, in the region immediately outside the horizon the tidal forces are so powerful that they can definitively separate each particle from its corresponding antiparticle to convert them into real particles or antiparticles. Particles/antiparticles that are closer to the horizon are absorbed, while those further away are free to travel to infinity, where they are detected by a distant observer as Hawking radiation. Since energy must be conserved, the absorbed particles have negative energy and the emitted particles have positive energy. This leads to a gradual reduction in the mass of the BH [9,10]. As Hawking showed, the radiation produced has a black body spectrum with an absolute temperature given by [10–12]:

\begin{equation}
T_{H} = \frac{hc^{3}}{16\pi^{2} kGM_{BH}},
\end{equation}

where $h = 6.63 \times 10^{-34} J\cdot s$ is the Planck constant and $k = 1.38 \times 10^{-23} J\cdot K^{-1}$ is the Boltzmann constant. An important feature of this equation is that $T \propto M_{BH}^{-1}$, which implies that $T_{H}$ is only significant for small and not very massive BHs. This is the Hawking temperature equation that we will heuristically derive using three different ways in the following sections.\\

Finally, it is important to emphasize that the argument developed in the two previous paragraphs shows us that the energy of Hawking radiation depends directly on the intensity of gravity (tidal forces) on the horizon. This point is explained in detail in [5]. On the other hand, according to Newton's law of gravitation, the gravity of a black hole is $g = GM_{BH}/r^{2}$; if we take $r = R_{S}$ we obtain: $g = c^{4}/4GM_{BH}$. Therefore, the energy of the radiation depends inversely on $M_{BH}$. But the energy of radiation also depends inversely on the wavelength. As a result, the wavelength $\lambda$ is proportional to $M_{BH}$ and $R_{S}$ (Eq. (1)). This conclusion is important because the derivations in the next section are based on it.

\section{First way: Wien's displacement law}

This derivation is a variant of a result presented in our previous work [4]. According to Wien's displacement law, the absolute temperature $T$ of a black body and the wavelength $\lambda_{max}$ where the emission maximum occurs are related as [13]:

\begin{equation}
\lambda_{max} T = \frac{hc}{5k}.
\end{equation}

But we know that the wavelength is proportional to the Schwarzschild radius, so that:

\begin{equation}
\lambda_{max} \sim R_{S} = \frac{2GM_{BH}}{c^{2}}.
\end{equation}

By introducing this relation into Eq. (3) and solving for $T$, we obtain a result that only differs from Eq. (2) in the dimensionless constants\footnote{An interesting aspect of this derivation is that if we proceed in the reverse direction and consider $T_{H}$ as given, we can show that $\lambda \sim R_{S}$. Indeed, it is easy to verify that Eq. (2) can be written as $T_{H} = hc/8\pi^{2} kR_{S}$. From this equality, if in Eq. (3) we take $T = T_{H}$, then $\lambda_{max} = (8\pi^{2} /5)R_{S} \cong 16R_{S} \sim R_{S}$. This means that the largest and most massive BHs emit Hawking radiation at longer wavelengths and therefore their Hawking temperature is comparatively lower, which is consistent with the fact that $T_{H} \propto M_{BH}^{-1}$.}:

\begin{equation}
T \sim \frac{hc^{3}}{10kGM_{BH}}.
\end{equation}

\section{Second way: Planck’s law}

Let us consider a photon of Hawking radiation of energy $E$, frequency $\nu$ and wavelength $\lambda$. According to Planck's law, we have [13]:

\begin{equation}
E = h\nu = \frac{hc}{\lambda}.
\end{equation}

But $\lambda \sim R_{S}$, so that: 

\begin{equation}
E \sim \frac{hc}{R_{S}} = \frac{hc^{3}}{2GM_{BH}}.
\end{equation}

Since Hawking radiation is thermal, we can take $E = kT$ [5], again obtaining a result that only differs from Eq. (2) in the dimensionless constants:

\begin{equation}
T \sim \frac{hc^{3}}{2kGM_{BH}}.
\end{equation}

\section{Third way: Uncertainty principle}

According to the uncertainty principle, the time between the appearance and the disappearance of a pair of virtual particles is [13]:

\begin{equation}
\Delta t \sim \frac{h}{4\pi \Delta E} = \frac{h}{4\pi E}
\end{equation}

where $\Delta E = E$ is the energy of the particles (for more details on the relations between virtual particles, uncertainty principle and black holes, see [6]). The maximum distance that the particles can then travel before disappearing is $c\Delta t \sim ch/4\pi E$. In order for tidal forces to convert virtual particles into real ones, they must separate them by a distance that cannot be less than $c\Delta t$. If we assume that $c\Delta t \sim \lambda$ then $c\Delta t \sim R_{S}$ so that:

\begin{equation}
\frac{ch}{4\pi E} \sim \frac{2GM_{BH}}{c^{2}}.
\end{equation}

Since $E = kT$, once again we obtain a result that only differs from Eq. (2) in the dimensionless constants:

\begin{equation}
T \sim \frac{hc^{3}}{8\pi kGM_{BH}}.
\end{equation}

\section{Final comments}

Equations (5), (8) and (11) only differ from Hawking's equation in dimensionless factors. However, in a heuristic derivation, these constants are of little importance. What we must highlight is the simplicity of the derivations, as well as the diversity of paths used to carry them out, which contributes to a broader and deeper understanding of the great Hawking’s findings.

\section*{Acknowledgments}
I would like to thank to Daniela Balieiro for their valuable comments in the writing of this paper. 

\section*{References}

[1] J. Pinochet, Hawking for beginners: a dimensional analysis activity to perform in the classroom, Phys. Educ. 55 (2020) 045018.

\vspace{2mm}

[2] J. Pinochet, Five misconceptions about black holes, Phys. Educ. 54 (2019) 55003.

\vspace{2mm}

[3] J. Pinochet, Stephen Hawking y los Agujeros Negros Cuánticos, Rev. Mex. Fís. E. 65 (2019) 182–190.

\vspace{2mm}

[4] J. Pinochet, “Black holes ain’t so black”: An introduction to the great discoveries of Stephen Hawking, Phys. Educ. 54 (2019) 035014.

\vspace{2mm}

[5] J. Pinochet, The Hawking temperature, the uncertainty principle and quantum black holes, Phys. Educ. 53 (2018) 065004.

\vspace{2mm}

[6] J. Pinochet, Hawking temperature: an elementary approach based on Newtonian mechanics and quantum theory, Phys. Educ. 51 (2016) 015010.

\vspace{2mm}

[7] J.P. Luminet, Black Holes, Cambridge University Press, Cambridge, 1999.

\vspace{2mm}

[8] V.P. Frolov, A. Zelnikov, Introduction to Black Hole Physics, Oxford University Press, Oxford, 2011.

\vspace{2mm}

[9] S.W. Hawking, A brief history of time, Bantam Books, New York, 1998.

\vspace{2mm}

[10] S.W. Hawking, The Universe in a Nutshell, Bantam Books, New York, 2001.

\vspace{2mm}

[11] S.W. Hawking, Black Hole explosions?, Nature. 248 (1974) 30–31.

\vspace{2mm}

[12] S.W. Hawking, Particle creation by black holes, Communications in Mathematical 
Physics. 43 (1975) 199–220.

\vspace{2mm}

[13] P.A. Tipler, R.A. Llewellyn, Modern Physics, 6th ed., W. H. Freeman and Company, New York, 2012.

\end{document}